\documentclass[twocolumn,showpacs,preprintnumbers,amsmath,amssymb]{revtex4-1}

\usepackage{graphicx,color}

\usepackage{dcolumn}

\usepackage{bm}

\begin{document}

\preprint{}

\title{A Quantal, Partially Ordered Electron Structure as a\\
       Basis for a $\gamma$ Free Electron Laser ($\gamma$-FEL).}

\author{D.~Habs$^{1,2}$, M.M.~G\"unther$^2$, S.~Karsch$^1$, P.G. Thirolf$^1$ and M.~Jentschel$^3$}

\affiliation{
$^1$Ludwig-Maximilians-Universit\"at M\"unchen, D-85748 Garching, Germany \\
$^2$ Max-Planck-Institut f\"ur Quantenoptik, D-85748 Garching, Germany \\
$^3$ Institut Laue-Langevin, F-38042 Grenoble, France}

\date{\today}

\begin{abstract}
When a rather cold electron bunch is transported during laser bubble acceleration in a strongly focusing plasma channel with typical forces of 100 GeV/m, it will form partially ordered long electron cylinders due to the relativistically longitudinal reduced repulsion between electrons, resulting in a long-range pair correlation function, when reaching energies in the laboratory above 0.5 GeV. During Compton back-scattering with a second laser, injected opposite to the electron bunch, the electron bunch will be further modulated with micro bunches and due to its ordered structure will reflect coherently, M\"ossbauer-like, resulting in a $\gamma$ free electron laser ($\gamma$-FEL). Increasing the relativistic $\gamma$ factor, the quantal regime becomes more dominant. We discuss the scaling laws with $\gamma$.
\end{abstract}


\maketitle

$\gamma$-ray lasers had been a longterm dream of laser physicists, trying toextend the laser wavelength into the $\gamma$ regime \cite{baldwin81,karyagin02}. In the last century the idea was to excite, e.g. with intense neutron pulses, shorter-lived nuclear isomers with M\"ossbauer transitions. However, the laser gain factors were too small in single-pass operation to obtain reasonable $\gamma$ beams. On the other hand, after FEL's~\cite{saldin00,schmueser08} reached shorter and shorter wavelengths, it was shown that the FEL principle for standard setups cannot be extended beyond $\approx 100$ keV, because for smaller wavelengths the thermal noise prevents stable micro-bunching of the electrons \cite{saldin04}. We will show in this publication that the very strong focusing in the plasma channel of laser-plasma 'bubble' acceleration \cite{pukhov02} enforces a very small bunch of electrons, which move extremely close together, due to the Lorentz transformation of the electron field into a flat, pancake-like distribution. The squeezing increases with electron energy that much, that a quantal system with partially degenerate Dirac-Fermi statistics occurs, where ordering prevails at higher temperatures. Thus FEL operation stops for classical electron beams beyond the 100 keV range, however, a new window of FEL operation opens up in the MeV $\gamma$ range due to the very strongly focusing plasma channel. The recoilless M\"ossbauer-like coherent emission results in a smooth emission without kicking individual electrons out of the ensemble.

\begin{figure} 
  \begin{center}
   \begin{tabular}{c}
   \includegraphics[height=6.0cm]{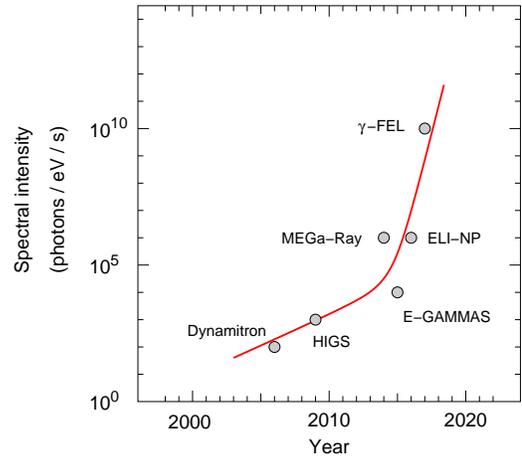}
   \end{tabular}
   \end{center}
   \caption[spec] 
  {\label{fig:spec} 
Spectral intensity of $\gamma$ beams. Following bremsstahlung beams like from the Dynamitron~\cite{beil80}, different types of Compton back-scattering sources are being developed. $\gamma$ FEL's may provide a large increase in spectral intensity.}
\end{figure}

Similar to X rays, where the 3. generation synchrotron light sources and now X-ray FEL's opened new worlds of application, we expect for $\gamma$ rays a similar development from Compton backscattering sources like HI$\gamma$S ~\cite{weller09}, MeGa-Ray~\cite{barty11}, E-GAMMAS~\cite{serafini12} or ELI-NP~\cite{ELI-NP11}, which one might call laser undulator sources, to significantly higher-quality $\gamma$-FEL beams. In Fig.~\ref{fig:spec} the observed or expected spectral intensities of $\gamma$ beams are shown as a function of their year of realization. 

A second important ingredient is short wave length optics. In 1996 A.~Snigirev et al.~\cite{snigirev96} realized refractive X-ray optics via Rayleigh scattering and allowed to optimize synchrotron beam parameters to experimental requirements \cite{nielsen01}. We recently found that it is possible to realize refractive optics for $\gamma$ rays based on Delbr\"uck scattering \cite{habs12}. We are expecting that such optics, based on Ge as a material, and in combination with double crystal monochromators \cite{kessler01}, to allow a very efficient extraction of a seed beam for a further FEL. Similar to X rays, we can develop many new applications for $\gamma$ rays, which we call nuclear photonics, addressing nuclei, not atoms, where the strong increase in $\gamma$ spectral intensity is essential. 

\begin{figure} 
   \begin{center}
   \begin{tabular}{c}
   \includegraphics[height=4cm]{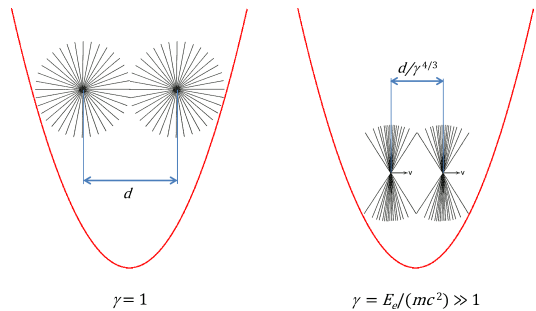}
   \end{tabular}
   \end{center}
   \caption[1d] 
   { \label{fig:1dcrystal} 
Coulomb fields between two electrons at rest ($\gamma=1$) and two relativistic electrons moving with a large $\gamma$ inside a harmonic static or co-moving trapping potential.}
  \end{figure}

In 2002 A.~Pukhov et al.\cite{pukhov02} proposed the laser-plasma ``bubble'' acceleration scheme, where a laser field expels in a small bubble all electrons from a plasma, forming a highly charged positve cavity behind the laser pulse, surrounded by the expelled electrons flowing around the wall of the bubble. From the experimental findings of laser bubble acceleration \cite{malka12} we know that typical focusing fields of several 100 GV/m occur in the plasma channel, where injected electrons have been accelerated up to 1 GeV \cite{leemans06}. Since the laser group velocity during bubble acceleration corresponds to a relativistic $\gamma\approx 10-20$ and is much smaller than the typical $\gamma=E_e/mc^2\approx 1000$ of the electrons with a final energy $E_e\approx 500$ MeV and an electron rest mass $mc^2$, we have to consider during acceleration a dynamical approach of the injected electron bunch towards
the center of the bubble, i.e. a slow approach towards the equilibrium configuration. In this paper we describe the final state of this dynamic process, which we expect to be very close to the equilibrium configuration. 
Transversely we face a harmonic focusing potential, where we assume rather cold electrons exhibiting basically no transverse velocity, or the betratron oscillations have been damped. If we consider electrons first at rest in such a strong harmonic trap, due to the repelling Coulomb force $\propto e^2/r^2$ and the focusing trapping force $F= k\cdot r$ with $k=100 GeV/m^2$, we obtain for cold charged particles a typical crystal spacing $d$~\cite{habs04,baus80,ichimaru82} with
\begin{equation}
  \label{eq:sherical}
 d=(\frac{e^2}{k})^{1/3}
\end{equation}

and a typical spacing  $d=1000\AA$. We have a one-component plasma (OCP)of electrons with a neutralizing ion background. If we now look for electrons, which are moving with a relativistic $\gamma=E_e/mc^2$, we get for typical values $\gamma=10^3$ the well-known pancake-like flattening of the electron field distribution, where the field strength is longitudinally reduced by a factor of $\gamma^{-2}$ and is transversely increased by a factor of $\gamma$ \cite{jackson72}. However, we now study two electrons in the harmonic focusing potential with both field distributions flattened as shown in Fig. \ref{fig:1dcrystal}. Thus, for two electrons placed along the velocity direction, we obtain a new distance $d_{long}$, which for an electron beam with $\gamma=1000$ results in a rather small distance $d_{long}=0.1 \AA$ with the surprising $\gamma^{-4/3}$ scaling:

\begin{equation}
  \label{eq:long}
 d_{long}=d/(\gamma)^{4/3}.
\end{equation}

For the central focusing potential the redistribution of the field lines of the electron with $\gamma$ is averaged out, while for the two electrons moving in velocity direction $\vec{v}$ only the field strength along their connecting line is important. On the other hand, if we place two electrons side by side, we get a $d_{trans}$ increased by $\gamma^{2/3}$, because of the two equivalent transverse directions compensating the longitudinal effect:

\begin{equation}
  \label{eq:trans}
 d_{trans}=d \cdot \gamma^{2/3}.
\end{equation}

We obtain a typical value of $d_{trans}=10^5 \AA= 10 \mu.$ These very different distances result in a {\it reordering} of crystal structures in the harmonic potential. The potential energy for the trapping potential in the longitudinal direction is a factor of $\gamma^2 =10^6$ less than for an arrangement normal to the velocity direction. Thus, even if we have a typical number of $N_e\approx 10^8$ electrons in the bunch, they all will arrange in $\approx 2-3$ cylinders with a diameter of $\approx 20 \mu$m and a length of also 20 $\mu$m in the laboratory system. We have $\approx 10^6$ lattice planes behind each other with a lattice spacingof 0.1 $\AA$ in the laboratory frame.  

Let us now consider the additional quantal repulsion of the fermionic electrons. From the Heisenberg uncertainty principle we know for the three space directions (i=x,y,z) at rest

\begin{equation}
  \label{eq:heisenberg}
     l_i\cdot p_i =h=  (1/c) (1.2 keV\cdot 100\AA)
\end{equation} 

since the longitudinal lattice spacing in the inner rest frame is a factor of $10^3$ smaller than the transverse one, we predominantly have to consider the longitudinal direction. However, we now consider the full coherent length of $10^6\cdot 100\AA$ of the crystal with a length of 10 mm in the rest frame and obtain a quantal energy of about 10 meV. With $\gamma=1000$ we find in the laboratory frame an energy of about 10 eV due to confinement in the lowest level, but we populate up to several MeV for a bunch of $\approx 20\mu m$ and 100 GeV/m in the quantal box and have the additional energy of 500 MeV due to the acceleration of the electron rest mass. At the outer egde of the crystal we find a Coulomb energy up to a few MeV, corresponding in the center of the crystal to a filling with about 20 electrons transversely and a few times $10^6$ electrons longitudinally. 

\begin{figure*} [t]
   \begin{center}
   \begin{tabular}{c}
   \includegraphics[height=4.9cm]{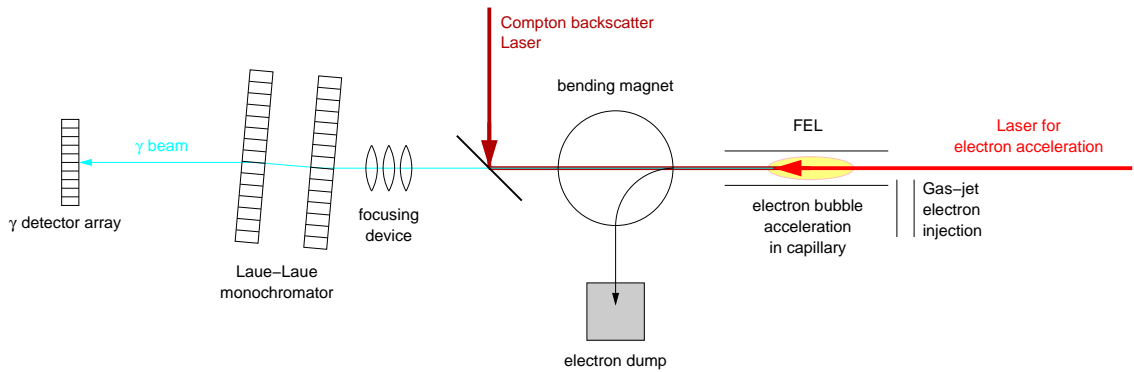}
   \end{tabular}
   \end{center}
   \caption[fel] 
   { \label{fig:fel}
   Experimental setup to measure the micro bunching inside the laser-plasma bubble via backscattered $\gamma$ quanta. With an efficient $\gamma$ monochromator using Delbr\"uck scattering and perfect germanium crystals, we can monochromatize between $\Delta E/E = 10^{-3}-10^{-6}$ and afterwards accurately measure the number of $\gamma$ quanta with their angular distribution, e.g. via $LaBr_3$ detectors at very low gain. Here we will use a new method to measure $\approx 10^8 \gamma$ quanta piling up within e.g. a few ps with high energy resolution by filtering with a double-crystal spectrometer.}
   \end{figure*}

The transversely increased Coulomb repulsion by $\gamma$ prohibits more transverse cylindrical layers for the lowest energy configuration. The typical Coulomb energy of about 10 electrons between longitudinal crystal layers is about 0.1 MeV and will preserve this structure also for the quantal system. On the other hand, the $10^7-10^8$ electrons will form a large quantal, partially degenerate system, like a very large atom. The large transition energies up to MeV and the large dipole moments with respect to the positive background will result in very fast cooling transitions below femtoseconds. This is only a qualitative picture: when increasing the relativistic $\gamma$, the number of cylindrical shells is reduced, on the other hand a larger number of electrons will condense into higher energy levels in the central part of the structure. The Fermi statistics results in reduced thermal fluctuations compared to the classical system. Thus we presently have only a simplified description, where more detailed studies in 3D and temperature will be added in a separate paper, looking also for the melting of the crystalline structure as a function of temperature and the cooling cascades.         

Also for the Compton back-scattered laser light of the FEL it is decisive that the velocity $v$ with $v/c\approx 1-1/(2\gamma^2)$ of the focussing and accelerating 'bubble' potential moves with $\gamma=10-20$, that is close to the velocity of light $c$ and similar to the final velocity of the electron bunch with $\gamma=1000$. Thus one can estimate the length $d_{crystal}\approx 500\mu$m in the laboratory frame, where the equilibrium configuration of the crystal stays together. Thus we have a long interaction time with the back-scatter laser, a long cooling time and the detailed acceleration of the electron bunch can be fine tuned.

The main idea of a free-electron laser (FEL) is that the produced high-energy photon beam with wavelength $\lambda_{\gamma}$ acts back onto the electron bunch by dispersion, inducing an equally spaced micro-bunching with distance $\lambda_{\gamma}$, so that all backscattered amplitudes add up coherently. In a first-order description, one introduces a Pierce or FEL parameter $\rho_{1d}$ to describe the high-gain FEL in the one-dimensional idealized case

\begin{equation}
   \rho_{1d}=\frac{1}{2\gamma}[\frac{I}{I_A}
(\frac{\lambda_u a_u}{2\pi\sigma_x})^2]^{1/3}.
\end{equation}

Here, I is the beam current, $I_A$=17 kA is the Alfv$\acute{e}$n current, $\sigma_x$ is the beam size in propagation direction, $a_u$ the undulator or laser intensity parameter and $\lambda_u$ is the wavelength of the stationary undulator, where we have for an optical laser undulator $\lambda_{Lu}= 2\cdot \lambda_{u}$.

For a typical bunch charge of 100~pC, a bunch length $\sigma_x= 10 \mu m$ or a duration of about 10 fs we obtain a current of I~=~10 kA. For a typical $\lambda_{Lu}= 1\mu$m , $a_u=1$ and $\gamma=10^3$ we obtain for $\rho_{1d}\approx 10^{-4}$. On the other hand, an energy spread of about 5 MeV is expected outside the laser-plasma bubble, where the additional Coulomb expansion has occurred after switching off the local focusing potential. Inside the bubble, the electron bunch will be much colder with a typical spacing of the quantum levels of 10 eV and the FEL conditionfor the electron energy spread

\begin{equation}
\frac{\Delta \gamma}{\gamma}\le \rho_{1d}
\end{equation}

will probably be fulfilled with an energy spread inside the bubble below about 50 keV. Since $I\propto \gamma$ and $\sigma_x\propto \gamma$, the Pierce parameter $\rho_{1d}$ will stay constantand since $\Delta \gamma/\gamma$ should improve with increasing $\gamma$, the relation should be better fulfilled for larger $\gamma$ values.  
The ideal gain length for micro-bunching is

\begin{equation}
L_{gain.ideal}=  \frac{\lambda_u}{4\pi \sqrt{3} \rho_{1d}}.
\end{equation}

Thus we obtain $L_{gain.ideal}\approx 50\mu m$. Typically eight gain lengths are required to reach saturation, and for the much larger length of the Compton back-scattered laser we should reach full FEL operation. If we have an FEL with a $\gamma$ wave length of about $1/100 \AA$ and a natural lattice spacing of the crystal of $1/10 \AA$, we can fulfill the Bragg reflection for about $n=\frac{2 d_{long}}{\lambda_{\gamma}}\approx 20$. Thus we have not a SASE-type FEL, but start from a coarse natural micro-bunching, which then is enhanced by the FEL bunching operation. However, if $10^8$ electrons  experience back-scattering the energy $E_e$ of the electron beam is reduced by $E_{\gamma}$, typically 1 MeV. Thus it will be essential to fine-tune the final acceleration with $MeV/\mu m$ in such a way, that the energy loss by back-scattering is compensated and each electron will back-scatter during the length $d_{crystal}$ many times. Therefore the intensity profile of the back-scattering laser is part of the fine-tuning. In this way the spectral intensity of the $\gamma$ beam can be strongly increasedto e.g. $10^{10}/(eVs)$. Moreover, we have to consider an FEL in the quantum regime, where a new quantum Pierce parameter has been defined:

\begin{equation}
\bar{\rho}=  \rho_{1d}\cdot(\frac{E_e}{E_{\gamma}}).
\end{equation}

For $\bar{\rho}\le 1$  the quantum FEL theory should be applied \cite{bonifacio08}. The quantum FEL has the interesting property that it results in a quantum purification of the FEL spectrum due to the large cooperation length, which in our treatment equals the bunch length. The M\"ossbauer-like recoil energy shift of the $\gamma$ rays is very small and we obtain a much more coherent $\gamma$ beam.

In the experiments we aim at reaching the coldest electron bunches possible \cite{malka12}. This not only will require very small transverse components, but it also will require a rather smooth high-contrast laser pulse, so that the trailing part of the short laser pulse will not lead to strong transverse heating of the electron bunch. We certainly do not intend a continuous injection resulting in rather long bunches. More narrow bunches could be achieved by a ``down ramped'' injection \cite{veisz10,malka12}, however, extending this to higher energies like 1 GeV, requires to find a combined solution of ``down ramping'' and a long plasma channel, which has not been realized at present, but probably can be found.

The essential new idea is to use high resolution $\gamma$ spectroscopy to study the electron bunch properties {\it inside} the focusing plasma channel. By monochromatization we can select with the new much larger Darwin width using the Delbr\"uck scattering a relative spectroscopic resolution ranging between $10^{-3}$ and $10^{-6}$ with a rather large efficiency of $\approx 20\%$. Thus we can filter simultaneously a number of quanta as large as e.g. $10^9$ without being hampered by pile-up problems. In a second step, we can measure the number of $\gamma$'s with much lower gain, determining the photon number in a certain angular range. We want to use an array of $LaBr_3$ scintillation detectors, making use of their fast temporal response and their high energy resolution capabilities. Thus we can measure the full forward angular distribution within a narrow spectral range. We can see from the reflected spectrum if a coherent reflection of all electrons occurs or if individual electrons reflect the laser photon and suffer a much larger recoil energy loss. Comparing the deduced electron spectrum, calculated  from the $\gamma$ spectrum with the Compton reflection, to the electron spectrum measured in an electron spectrometer after leaving the plasma channel, allows to verify the quantal collective picture as a function of the electron beam energy. We can easily detect an expected strong coherent enhancement from the number of emitted $\gamma$ quanta. Thus we can optimize the FEL operation. The experimental setup can be used for diagnostics, but also for working with very high production rates.

Once we observe that a coherent enhancement is obtained, in a second step we could seed an FEL with a further monochromatized $\gamma$ beam and reach in a second step an even larger enhancement of the spectral intensity by potentially some orders of magnitude. We can reach an energy resolution in the meV range for these M\"ossbauer-like transitions. If in the more long-term future, high-power laser systems with repetition rates around 1 kHz become available, even $10^3$-fold higher spectral intensities could be achieved.  

We acknowledge very helpful discussions with K.~Witte and M.~Gross.

\end{document}